\documentclass{iau} 
\usepackage{graphicx}

\newcommand\aj{AJ} 
\newcommand\apj{ApJ}

\newcommand\aap{A\&A} 
\newcommand\mnras{MNRAS} 
\newcommand\apjl{ApJ}     
 
\newcommand\nat{Nature} 
 
\newcommand\aapr{A\&AR}

\title[Stellar populations in Magellanic-Cloud clusters] 
{The {\it HST} survey of Magellanic-Cloud clusters and of their stellar populations}

\author[A.\,P.\,Milone]   
{A.\,P.\,Milone$^1$
}

\affiliation{$^1$ Research School of Astronomy \& Astrophysics, Australian National University, Mt Stromlo Observatory, via Cotter Rd, Weston, ACT 2611, Australia.\\ email: {\tt milone@mso.anu.edu.au}}

\pubyear{2015}
\volume{316}  
\jname{Formation, evolution, and survival of massive star clusters}
\editors{Corinne Charbonnel \&  Antonella Nota, eds.}
\begin{document}

\maketitle

\begin{abstract}
A large number of intermediate-age ($\sim$1-2-Gyr old) globular clusters (GCs) in the Large and the Small Magellanic Cloud (MC) exhibit either bimodal or extended main-sequence (MS) turn off and dual red clump (RC). Moreover, recent papers have shown that the MS of the young clusters NGC\,1844 and NGC\,1856 is either broadened or split. These features of the color-magnitude diagram (CMD) are not consistent with a single isochrone and suggest that star clusters in MCs have experienced a prolonged star formation, in close analogy with Milky-Way GCs with multiple stellar populations. As an alternative, stellar rotation or interacting binaries can be responsible of the CMD morphology. In the following I will summarize the observational scenario and provide constraints on the nature of the complex CMD of young and intermediate-age MC clusters from our ongoing photometric survey with the {\it Hubble Space Telescope} ({\it HST\,}).
\end{abstract}

\section{Introduction}
It is now widely accepted that the CMD of nearly all the old Galactic GCs (GGCs) is made of multiple sequences that can be identified along the entire diagram, from the bottom of the MS up to the sub-giant branch (SGB), the red-giant branch (RGB) and even the horizontal branch and the asymptotic giant branch (e.g.\,Milone et al.\,2012; Piotto et al.\,2015).
The multiple sequences correspond to distinct stellar populations with different helium and light-element abundance and their presence has suggested that GCs have experienced a complex star-formation history.

The discovery of either multimodal or extended MS turn-off (hereafter eMSTO) in several intermediate-age star clusters in both MCs (e.g.\,Bertelli et al.\,2003; Mackey \& Broby Nielsen\,2007; Glatt et al.\,2008) has suggested that multiple stellar populations could not be a peculiarity of old GGCs. 
 In our survey of MC star clusters, we have shown that the the eMSTO is a common feature among $\sim$1-2-Gyr old star clusters and has been detected in at least the 70\% of the analyzed objects (Milone et al.\,2009).

\section{The CMD of eMSTO clusters}
\begin{figure}[b]
\begin{center}
 \includegraphics[width=4.5in]{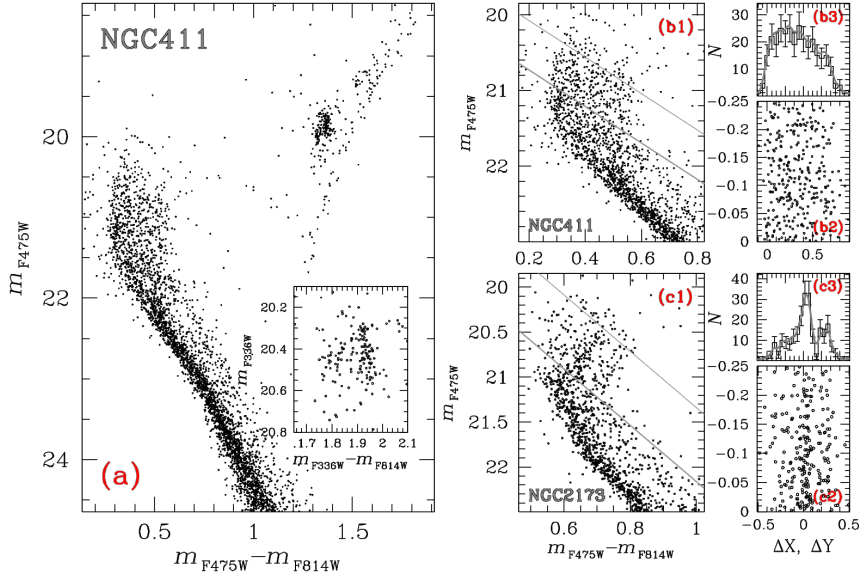}
 \caption{\textit{Panel (a):} $m_{\rm F475W}$ vs.\,$m_{\rm F475W}-m_{\rm F814W}$ CMD of the intermediate-age GC NGC\,411. The inset is a zoom of the $m_{\rm F336W}$ vs.\,$m_{\rm F336W}-m_{\rm F814W}$ around the RC. Panels b1 and c1 show the $m_{\rm F475W}$ vs.\,$m_{\rm F475W}-m_{\rm F814W}$ CMD for stars around the MSTO in NGC\,411 and NGC\,2173, respectively. The verticalized color and magnitude distribution for the MSTO stars between the two gray lines of panels a1 and b1 are plotted in panels a2 and b2, while panels a3 and b3 show the histograms of the $\Delta$X distribution along the direction, X, of the gray lines for these stars. The gray lines overimposed on each histogram are the corresponding kernel-density distributions.}
   \label{fig:cmd}
\end{center}
\end{figure}

An example of typical CMD of an intermediate-age star cluster is provided in Fig.~\ref{fig:cmd}a, where I have plotted $m_{\rm F475W}$ vs.\,$m_{\rm F475W}-m_{\rm F814W}$ for NGC\,411 in the SMC.  As mentioned above, the most striking feature is the MSTO, which is widely broadened in color and magnitude.
 Despite some authors have initially claimed that the eMSTO is due to photometric errors (see e.g.\,Mucciarelli et al.\,2007 for the case of NGC\,1783), the fact that the other CMD sequences, like the RGB and the lower MS are narrow and well defined, clearly demonstrates that the broadened MSTO is intrinsic and is not due neither to differential reddening nor to photometric errors.

The distribution of stars along the MSTO changes from one cluster to another. Some eMSTO clusters, like NGC\,411, exhibit a continuous distribution (see Fig.~\ref{fig:cmd}b1-b3). In other clusters there are two main MSTOs, with the faint MSTO hosting approximately 30-40\% of the total number of MSTO stars (see Fig.~\ref{fig:cmd}c1-c3 for the case of NGC\,2173). 
Another intriguing feature 
 is the RC, which  can be either broadened or bimodal (Girardi et al.\,2009) as shown in the inset of Fig.~\ref{fig:cmd}a for NGC\,411. 

The dual RC has been interpreted as due to two stellar groups:  one of them consisting of massive stars, which avoid e$^{-}$ degeneracy settling in their H-exhausted cores when He ignites. The other group is made of slightly less-massive stars which experience e$^{-}$ degeneracy before He ignition. This phenomenon is thought to be responsible for the higher brightness of the less-massive stars (Girardi et al.\,2009, 2013).

\subsection{The RGB}
\label{sec:rgb}
As discussed in the previous section, most GGCs host two or more stellar populations with different chemical composition. One of them has the same C, N, O, and He abundance as halo-field stars with the same metallicity, while the other population(s) is enhanced in helium and nitrogen and depleted in carbon and oxygen (see Gratton et al.\,2012; Piotto et al.\,2015 and references therein).
As a consequence of such a specific chemical pattern, most GGCs exhibit either broadened or multiple RGBs, when observed in appropriate CMDs involving the U band (like $U$ vs.\, $U-B$, Marino et al.\,2008).

 Recent papers, based on spectroscopy and {\it HST} photometry, have shown that N/He-rich stars of GGCs have larger $m_{\rm F336W}-m_{\rm F438W}$ color than N/He-poor stars with the same luminosity, but smaller $m_{\rm F438W}-m_{\rm F814W}$ color. The $C_{\rm F336W, F438W, F814W}=$($m_{\rm F336W}-m_{\rm F438W}$)$-$($m_{\rm F438W}-m_{\rm F814W}$)  index introduced by Milone et al.\,(2013a) thus maximizes the virtues of both $m_{\rm F336W}-m_{\rm F438W}$ and $m_{\rm F438W}-m_{\rm F814W}$, and is a powerful tool to identify multiple MSs and RGBs.

\begin{figure}[b]
\begin{center}
 \includegraphics[width=2.62in]{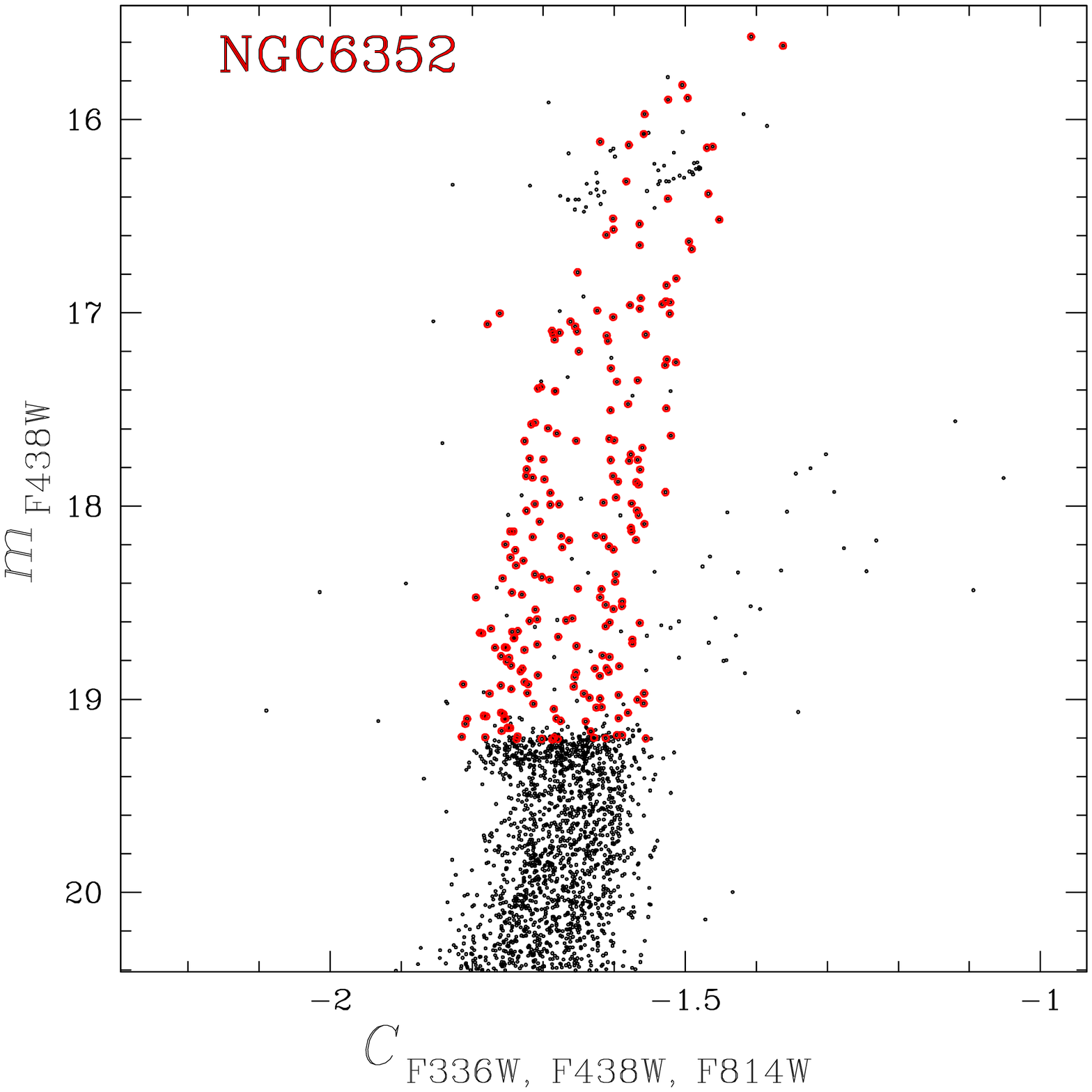}
 \includegraphics[width=2.62in]{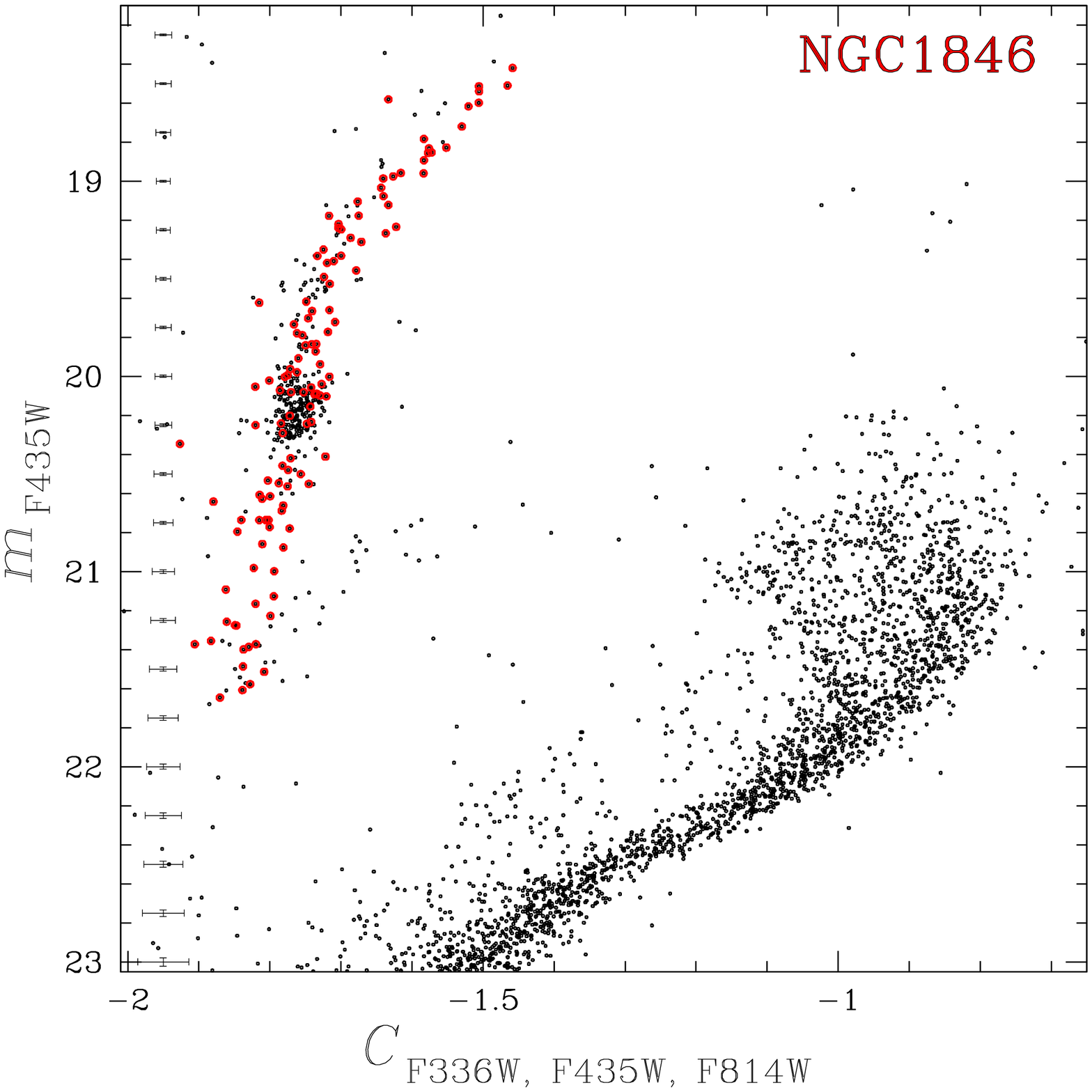}
 \caption{$m_{\rm F438W}$ vs.\,$C_{\rm F336W, F438W, F814W}$ diagram for the GGC NGC\,6352 (left) and $m_{\rm F435W}$ vs.\,$C_{\rm F336W, F435W, F814W}$ diagram for NGC\,1846 in the LMC (right). Red circles mark RGB stars.}
   \label{fig:rgb}
\end{center}
\end{figure}

As an example, the left panel of Fig.~\ref{fig:rgb} shows $m_{\rm F438W}$ vs.\,$C_{\rm F336W, F438W, F814W}$ for NGC\,6352, that is a metal-rich GGC ([Fe/H]=$-$0.64, Harris 1996, 2010 edition) with mass, $\mathcal{M}=1.2 \times 10^{5} \mathcal{M}_{\rm \odot}$ (Gnedin \& Ostriker\,1997). NGC\,6352 hosts two distinct stellar populations with different He, C, N, Na, and O, that populate the two distinct sequences detected along the entire CMD by Piotto et al.\,(2015) and Nardiello et al.\,(2015). The split RGB  of NGC\,6352, is clearly visible in the left-panel diagram of Fig.~\ref{fig:rgb} from Piotto et al.\,(2015), where I have marked RGB stars with red circles.

In the right panel of Fig.~\ref{fig:rgb}, I show $m_{\rm F435W}$ against $C_{\rm F336W, F435W, F814W}$ for the eMSTO LMC cluster NGC\,1846, which has similar mass ($\mathcal{M}=2.3 \times 10^{5}  \mathcal{M}_{\rm \odot}$, Goudfrooij et al.\,2014) and metallicity ([Fe/H]=$-$0.4, Milone et al.\,2009) as NGC\,6352. In this case the RGB width in the $C_{\rm F336W, F438W, F814W}$-index, is consistent with the broadening expected from photometric errors, thus demonstrating that star-to-star variations in light-elements and helium, if present, are small and can  not be detected from ultraviolet photometry of RGB stars.      
 This finding, corroborates the conclusion by Mucciarelli et al.\,(2014) who have analyzed spectra of eight bright red giants in the eMSTO cluster NGC\,1806 and found homogeneous Na and O content.

\section{Theoretical interpretation}
 Several authors have interpreted the eMSTO of intermediate-age MC clusters as due to a prolonged period of star formation with a duration of $\sim$100-500 (see Goudfrooij et al.\,2014 and references therein).
 An alternative interpretation has been provided by Bastian \& de Mink (2009) who have noticed that rotation significantly affects the stellar structure and the inclination angle between the observer and the star changes the effective temperature. They have thus suggested that a range of stellar rotation in MSTO stars can mimic an eMSTO in a cluster where a simple stellar population exists. 
 Unfortunately, as discussed by Yang et al.\,(2013) the evolution of rotating stars is strongly dependent on the stellar model and the eMSTO can be explained or not depending on the adopted models.
 As a further interpretation of the CMD of intermediate-age MC clusters, interacting binaries with mass transfer and merged binary systems have been also considered as possible responsible for the eMSTO (Yang et al.\,2011).

A lot of effort has been done by several groups to discriminate among the different scenarios. Most of their works are based on the comparison between theoretical models and observation of stars along the eMSTO, the RC, and the SGB. I refer to papers by Goudfrooij et al.\,(2015), Li et al.\,(2014), Bastian et al.\,(2015) Niederhofer et al.\,(2015) and references therein for critical discussion about age spread and rotation as responsible of the eMSTO. 

\section{The young clusters NGC\,1844 and NGC\,1856}
\begin{figure}[!htb]
\begin{center}
 \includegraphics[width=2.62in]{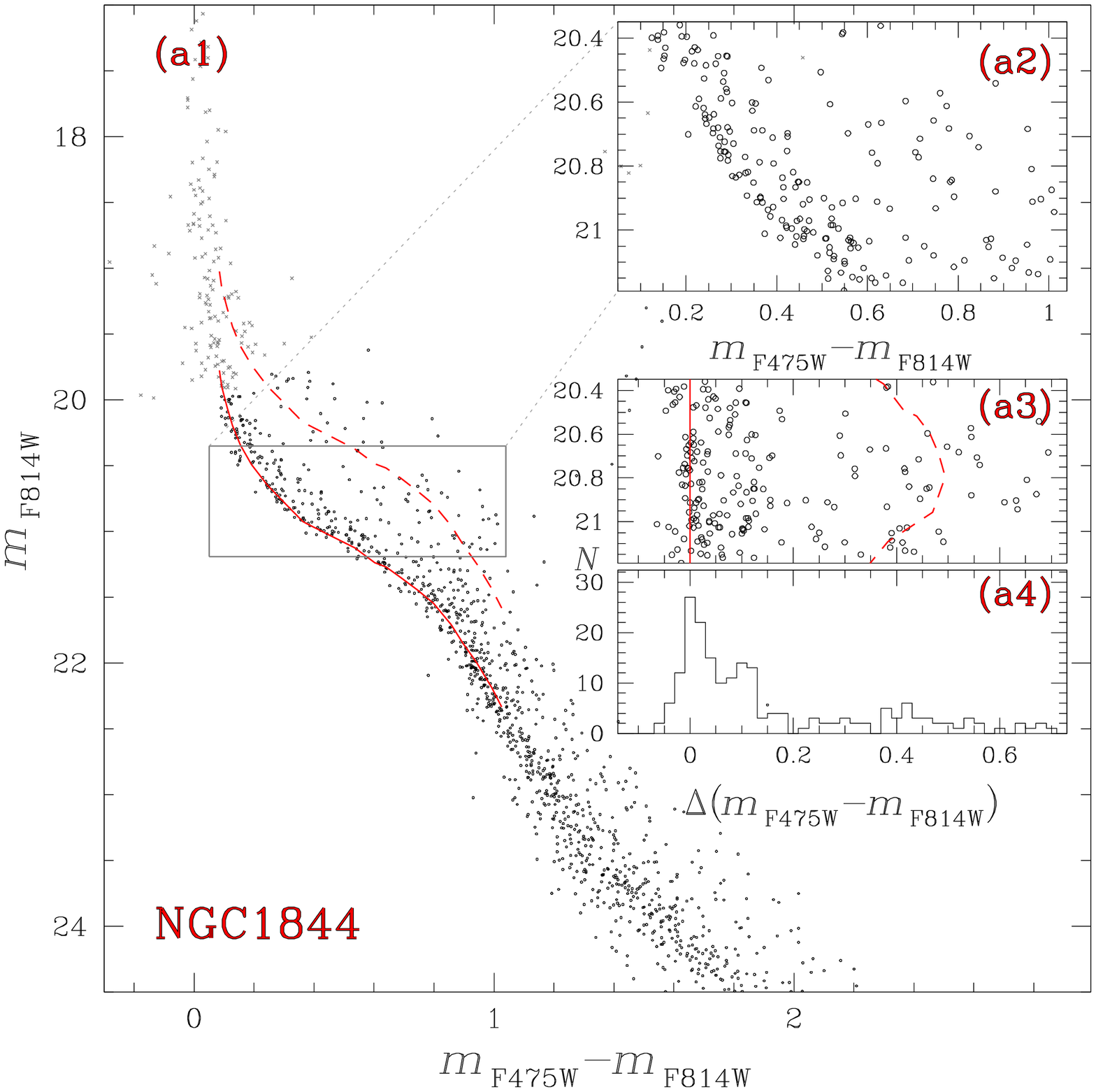}
 \includegraphics[width=2.62in]{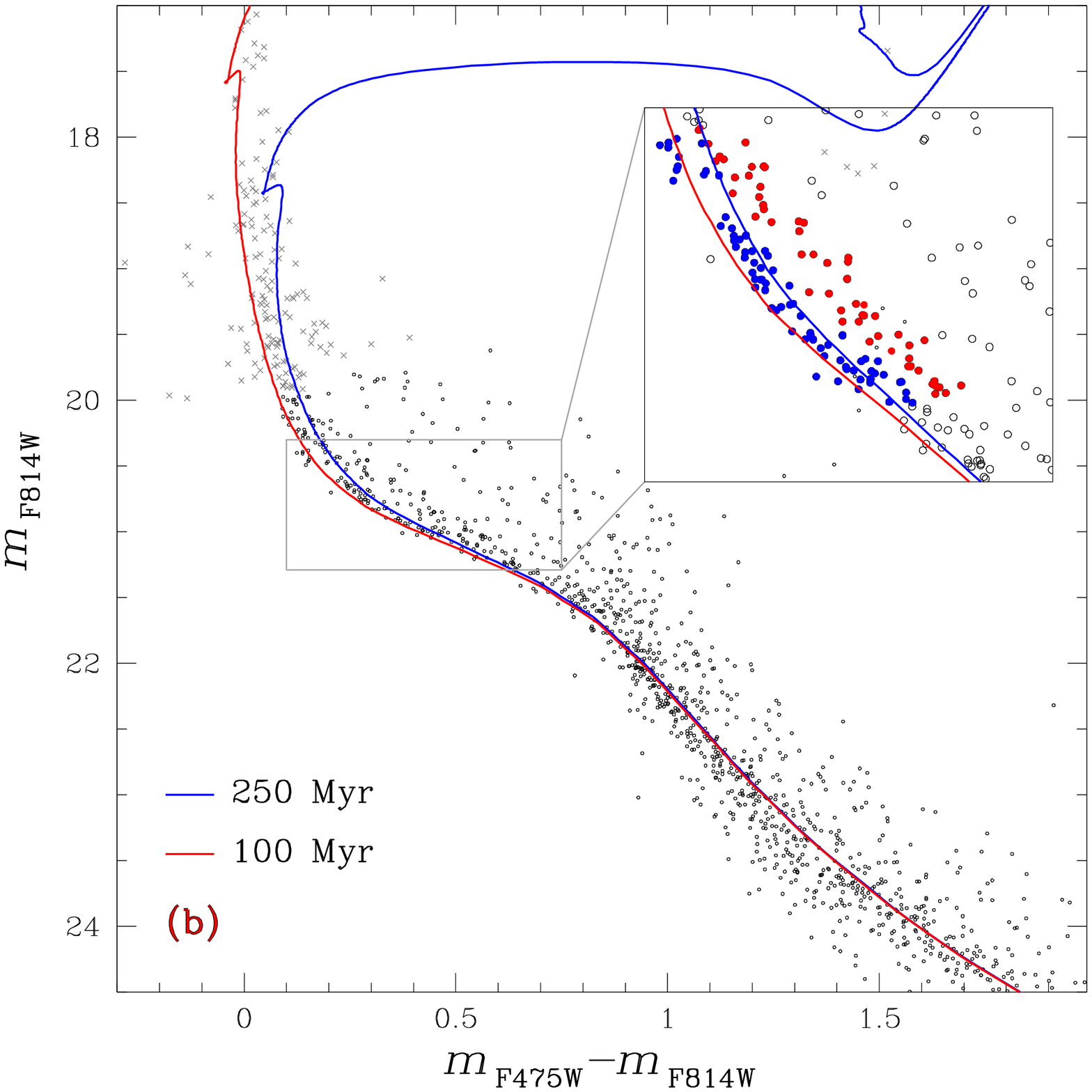}
 \caption{\textit{Left panels:} CMD of the LMC cluster NGC\,1844, with saturated stars colored gray. The red continuous and dashed lines mark the MS fiducial line, and the line of equal-mass binaries, respectively (panel a1). A zoom of the CMD region where the MS broadening is more-clearly visible is provided in panel a2. Panel a3 shows the verticalized $m_{\rm F475W}$ vs.\,$\Delta (m_{\rm F475W}-m_{\rm F814W})$ diagram for the MS stars plotted in panel a2, where $\Delta (m_{\rm F475W}-m_{\rm F814W})$ is obtained by subtracting from the $m_{\rm F475W}-m_{\rm F814W}$ color of each star, the corresponding color of the MS fiducial line at the same F475W magnitude. The histogram distribution of $\Delta (m_{\rm F475W}-m_{\rm F814W})$ is plotted in panel a4. 
\textit{Right panels:} Reproduction of the same CMD of panel a1 with overimposed BaSTI isochrones with metallicty Z=0.01 and ages of 100 and 250 Myr.  
 }
   \label{fig:ngc1844}
\end{center}
\end{figure}

New insight on the eMSTO phenomenon comes from young clusters.
 In this context the $\sim$150-Myr old NGC\,1844 in the LMC is exemplary. As shown in Fig.~\ref{fig:ngc1844} the upper MS of this cluster is widely spread in $m_{\rm F475W}-m_{\rm F814W}$ color, with some hints of MS split (Milone et al.\,2013b). The MS broadening is intrinsic and can not be due neither to field-star contamination, non-interacting binaries, photometric errors, or variation of the foreground reddening. 
 Unfortunately, the interpretation of the CMD of NGC\,1844 is still controversial.
  The comparison of the observations  with BaSTI isochrones (Pietrinferni et al.\,2004) with the same metallicity but different ages reveals that age difference alone can not be responsible for the broadened MS of NGC\,1844 (Fig.~\ref{fig:ngc1844}b).
 Similarly, on the basis of isochrones from the same database, Milone et al.\,(2013b) have excluded that the MS of NGC\,1844 is consistent with a range of stellar rotation, while isochrones with different overall C$+$N$+$O abundance could provide a better fit with the observed CMD. 

\begin{figure}[!htb]
\begin{center}
 \includegraphics[width=3.4in]{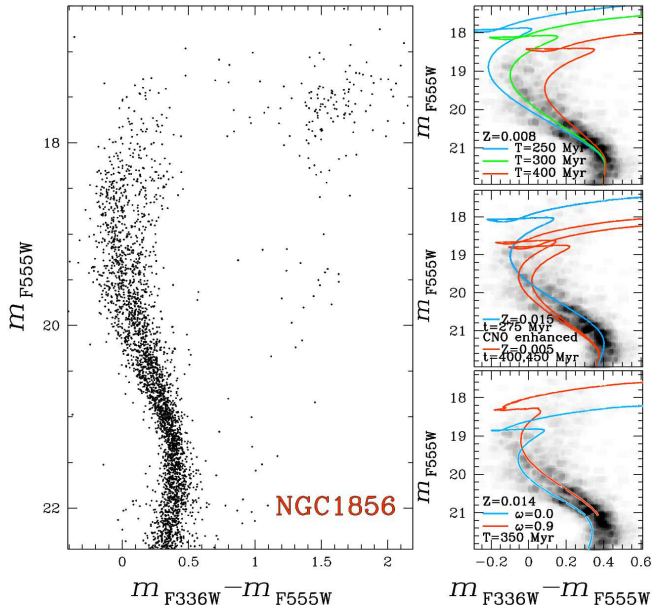}
 \caption{\textit{Left panel.} CMD of the LMC GC NGC\,1856 from WFC3/{\it HST} photometry (Milone et al.\,2015). \textit{Right panels.} Isochrones are compared with the Hess diagram of NGC\,1856. BaSTI isochrones with the same metallicity but different age are plotted in the upper panel, while the middle panel shows BaSTI isochrones with different age, [Fe/H] and C$+$N$+$O abundance. The lower panel compares two coeval isochrones from the Geneva database with no rotation (blue line), and with high-rotation (red line, D'Antona et al.\,2015).}
   \label{fig:ngc1856}
\end{center}
\end{figure}

Another intriguing GC is the $\sim$350 Myr-old NGC\,1856, which is the only cluster younger than $\sim$1 Gyr that exhibits an eMSTO (Milone et al.\,2015; Correnti et al.\,2015). In addition, Milone et al.\,(2015) have revealed that the MS of NGC\,1856 is bimodal, with the blue MS hosting about one third of the total number of MS stars. 
 As shown in Fig.~\ref{fig:ngc1856}, the split MS of NGC\,1856 is visible only at bright luminosity and the two MS merge together  approximately two F555W mag below the MSTO. It is not clear whether the blue MS evolves into the faint MSTO or if it is connected with the bluest part of the MS.
 Similarly to the MS, the RC of NGC\,1856 is bimodal.
In contrast, there is no evidence for either split or broadened MS in none of the analyzed intermediate-age GCs.

The interpretation of the CMD of NGC\,1856 is still debated. As shown in the upper-left panel of Fig.~\ref{fig:ngc1856}, where I have overimposed isochrones withe the same metallicity on the Hess diagram of NGC\,1856, if the cluster stars are chemically homogeneous, the eMSTO is consistent with a prolonged star-formation history of about 150 Myr. In this case, the blue MS would evolve into the brighter MSTO. In contrast, if the two MS merge together and the blue MS is connected with the faint MSTO, the CMD of NGC\,1856 is consistent with two stellar populations with different metallicity and age. In this case the blue MS would be enhanced in [Fe/H] and [(C$+$N$+$O)/Fe] and younger than the red MS (middle-left panel of Fig.~\ref{fig:ngc1856}).
 However, while metallicity variations have been detected in some GGCs (see Marino et al.\,2015 and references therein), as discussed in Sect.~\ref{sec:rgb}, there is no evidence for star-to-star light-element variations in intermediate-age MC clusters.

As an alternative, according to Geneva stellar models (Georgy et al.\,2014), the observations of NGC\,1856 are consistent with a coeval assembly of stars with a range of rotation. In this context, both the bright MSTO and the red MS of NGC\,1856 correspond to a population of rapidly-rotating stars, while the blue-MS and the faint MSTO are made of non-rotating stars (D'Antona et al.\,2015, see bottom-left panel of Fig.~\ref{fig:ngc1856}).  
\section{Conclusions}
I have used the photometry from our ongoing survey of MC cluster with {\it HST} to illustrate the main CMD features of intermediate-age and young clusters that are not consistent with a single isochrone. 
 These features include the eMSTO and the dual RC, which are commonly observed in most GCs with age of $\sim$1-2 Gyr, and the broadened and double MS, recently discovered in the young clusters NGC\,1844 and NGC\,1856. Moreover, I have shown that the RGB of NGC\,1846 shows no evidence for light-elements variations, in contrast with what observed in the majority of old GGCs. 

The presence of multiple generations of stars in intermediate and young MC clusters is still controversial.
If the eMSTO is due to age difference, these clusters would provide a unique angle to study the multiple-population phenomenon.  Indeed they could be similar to the old GGCs just few hundreds Myrs after their formation (e.\,g.\,Keller et al.\,2011). 
On the contrary, if only one population exists, the eMSTO star clusters would provide unique laboratories to study the physic of rotating stars or interacting binaries.

\end{document}